# Weather-Adaptive Multi-Step Forecasting of State of Polarization Changes in Aerial Fibers Using Wavelet Neural Networks

Khouloud Abdelli[1], Matteo Lonardi[2], Jurgen Gripp[3], Samuel Olsson[3], Fabien Boitier[4], and Patricia Layec[4]

[1] Nokia Bell Labs, Germany Khouloud.Abdelli@nokia.com
[2] Nokia, Vimercate, Italy, [3] Nokia, Murray Hill, NJ, 07974 USA, [4] Nokia Bell Labs, France

**Abstract** *We introduce a novel weather-adaptive approach for multi-step forecasting of multi-scale SOP changes in aerial fiber links. By harnessing the discrete wavelet transform and incorporating weather data, our approach improves forecasting accuracy by over 65% in RMSE and 63% in MAPE compared to baselines.* ©2024 The Author(s)

**Introduction**

Accurately forecasting state of polarization (SOP) changes is paramount for enhancing anomaly detection, prediction, and maintenance strategies, thereby ensuring the resilience of optical networks amidst dynamic environmental conditions. Yet, predicting SOP changes in aerial optical fibers presents a multifaceted challenge owing to their distinct attributes and the intricacies of environmental dynamics. These fibers are particularly susceptible to external influences, leading to abrupt or unforeseeable shifts in light polarization [1-3]. These fluctuations profoundly affect the performance and reliability of optical communication systems, often triggered by diverse factors such as atmospheric conditions, wind-induced vibrations, and other environmental variables. Rapid SOP changes often align with severe weather events like storms [4]. Mechanical vibrations affecting SOP span from milliseconds to minutes or longer. Wind-induced vibrations swiftly alter SOP within seconds, while seismic activity induces more gradual changes over minutes. Temperature variations induce gradual SOP drift over minutes to hours, while humidity fluctuations affect fiber properties, contributing to extended SOP changes.

We propose a novel weather-adaptive strategy for forecasting both rapid and gradual SOP variations. Our method strategically blends short-term wind-induced and long-term gradual SOP change prediction techniques to enhance accuracy. We leverage a multiresolution prediction framework based on the discrete wavelet transform, which allows us to seamlessly integrate wind dynamics, temperature variations, and humidity levels into the forecasting process. We validate our approach with field trial SOP data.

**Adaptive SOP Forecasting Framework**

Fig. 1 illustrates our proposed framework for multi-scale SOP change forecasting, which integrates environmental factors including wind, temperature, and humidity. Our approach combines short-term and long-term forecasting methods to capture the dynamics of SOP changes. The short-term forecasting method operates at the micro-scale, focusing on detecting rapid wind-induced fluctuations within time windows ranging from milliseconds to seconds. In contrast, the long-term forecasting method operates at the macro-scale, predicting gradual SOP changes over minutes. Historical SOP measurements and weather data are collected and pre-processed for both forecasting methods, with sampling rates tailored to each. Wind data is collected for the short-term forecasting method, providing real-time insights into rapid SOP fluctuations induced by wind dynamics. Meanwhile, temperature and humidity data are gathered for the long-term forecasting method, allowing for the analysis of gradual SOP changes influenced by environmental conditions over extended periods. To address inherent noise in SOP measurements from aerial fibers, denoising techniques such as exponential moving averages are applied. Next, the pre-processed data undergoes decomposition into approximations and details using the discrete wavelet transform (DWT) technique [5-8] before being fed into the corresponding forecasting model (short-term or long-term). In scenarios where windy conditions are forecasted for the upcoming window, the short-term micro-scale predictions are integrated into the minute-scale predictions. Short-term predictions are aggregated to form minute-scale predictions However, if such conditions are not anticipated, the minute-scale predictions rely solely on the long-term model outputs.

**Short-term SOP Change Prediction**

We integrate wind effects into the short-term SOP forecasting method, called "Windy Model" in the following. The "Windy Model" is a multivariate approach that utilizes historical and current wind gust and SOP change data to forecast future SOP changes. Initially, both the SOP change and wind gust data undergo decomposition using DWT with the db5 wavelet basis resulting in approximation coefficient $A_1$ and detail components $(D_1, D_2, D_3, D_4, D_5)$ for SOP change data, and approximation coefficient $A_1^w$ and $(D_1^w, D_2^w, D_3^w, D_4^w, D_5^w)$ for wind gust data. Through a multiresolution analysis using DWT, we evaluate the correlations between SOP change and wind gust time-series at various frequency levels. The components demonstrating the highest correlation are selected as inputs to the Artificial Neural Network (ANN) structure. Typically, the approximation coefficients exhibit the strongest correlation. Thus, both the approximation coefficients of SOP change and wind gust data serve as inputs to the ANN model to predict the next forecasted SOP change data's approximation coefficient. Meanwhile, the SOP change details are used solely to forecast the details of the next forecasted SOP change's DWT composition, as depicted in Fig. 2. Finally, the forecasted components are reconstructed using an Inverse DWT (IDWT) [6,7] to obtain the predicted SOP change data.

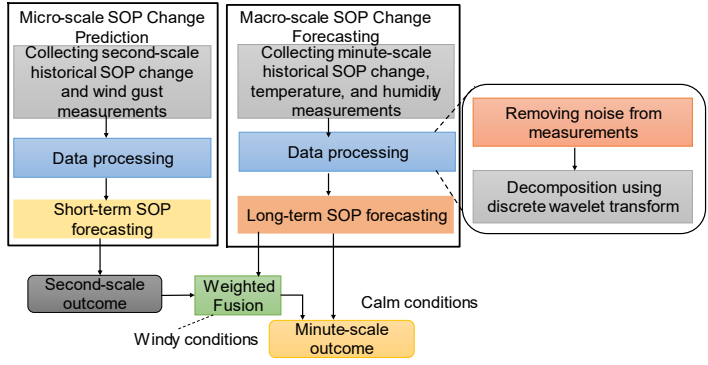
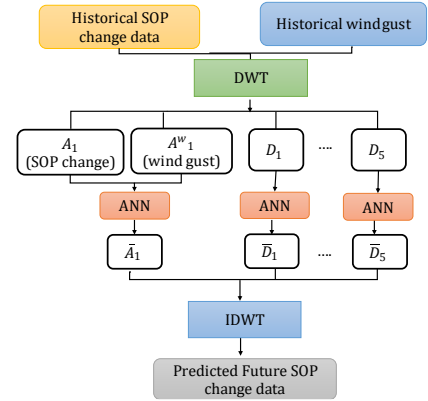

Fig. 1: Proposed approach for multi-scale SOP change forecasting.

Fig. 2: Structure of windy model.

**Long-term SOP Change Prediction**
The forecasting model for long-term SOP change forecasting integrates temperature and humidity data alongside SOP change data. Like the structure of the "Windy Model", our model retains a similar framework, with the adaptation of replacing wind gust data with humidity and temperature data. The approximation coefficients of the SOP change and temperature and humidity serve as an input to ANN model to predict the approximate coefficient of the forecasted SOP change data, whereas only the details of the SOP change data are fed into separate ANN models to predict the details of the DWT of the forecasted SOP change data.

**Field Trial Data**
Our framework is assessed using real-world SOP change data obtained from aerial optical fibers, coupled with weather data. The field trial (Fig. 3(a)) records SOP data along a bidirectional link spanning over 940 km between locations A and B. Raman amplifiers are used for compensating signal attenuation. SOP monitoring is integrated into a coherent transponder within this trial. Data collection occurred from March 18 to May 31, 2021, at a sampling rate of 1 second for short-term forecasting and 30 minutes for long-term forecasting. SOP changes are measured based on SOP rotation speed, derived from Stokes parameters. Denoising via exponential smoothing was applied to the collected data, resulting in mean values of 211 rad/s and 209 rad/s for short-term and long-term forecasting, with standard deviations of 42 and 35, respectively.

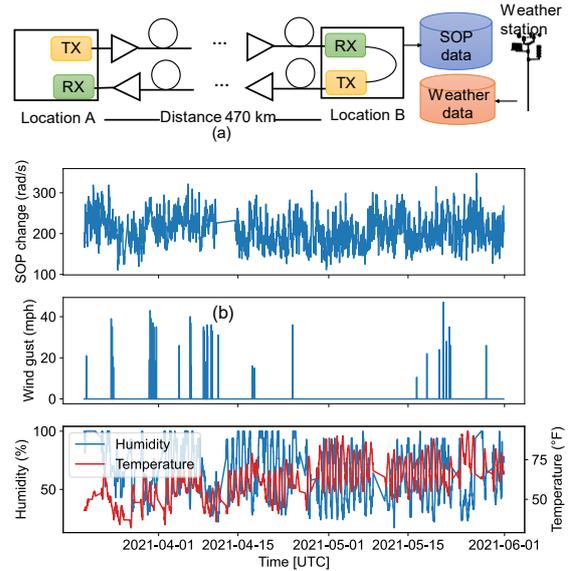

Fig. 3: (a) Field trial for SOP change data recording, (b) SOP change and weather data.

The weather data, sourced from a nearby weather station in B via Weather Underground [9], includes 14 features such as wind gust, temperature, and humidity. Aligning with SOP change data from March 18 to May 31, 2021, the weather reports were collected at 30-minute intervals. To match the 1-second intervals of the SOP data, interpolation technique was applied. Fig. 3 (b) illustrates the SOP change data collected at B, along with wind gust, humidity, and temperature data sampled at 30-minute intervals. The data, spanning from March 18 to May 26, 2021, was utilized for training the forecasting models, while the data from May 26 to May 31, 2021, served for model evaluation. The data was divided into fixed-length sequences, with a length of 36 for short-term forecasting (equivalent to 36s) and 48 for long-term forecasting (representing one day of historical SOP change). The short-term method predicts the next 12 SOP change values (12s ahead), while the long-term method forecasts SOP changes for the next 12 hours.

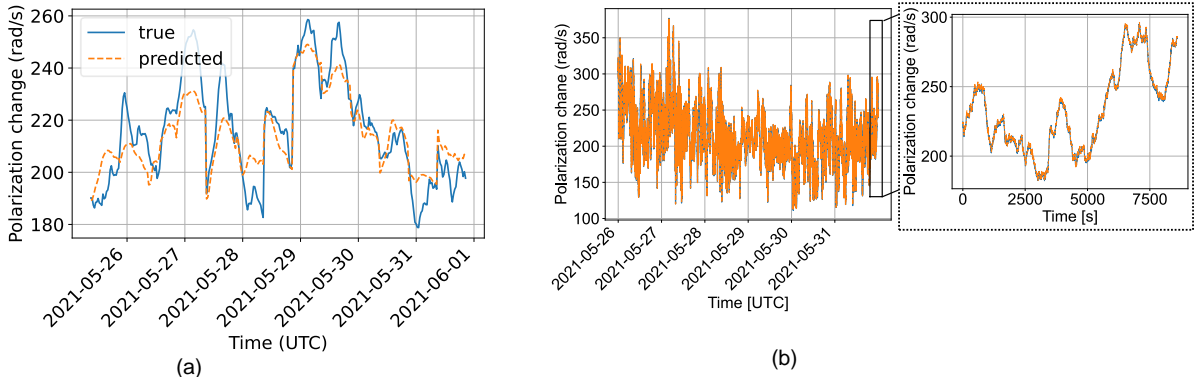

Fig. 4: Prediction results for: (a) long-term forecasting, (b) short-term forecasting (same legend as (a)).

**Results**

Our proposed Windy model for short-term forecasting has been evaluated using Root Mean Square Error (RMSE) and Mean Absolute Percentage Error (MAPE) metrics. It achieves a remarkably low RMSE of 0.89 rad/s and MAPE of 0.3%, marking improvements of 30.47% and 50% respectively compared to our approach when trained solely on SOP data without incorporating wind gust data, termed the "Calm Model", highlighting the importance of integrating wind data. We further benchmark our approach against other baselines, including ANN and the conventional moving average method. The ANN model was trained using historical measurements without undergoing wavelet decomposition. As depicted in Tab.1, our approach outperforms the other methods, demonstrating the lowest RMSE and MAPE values. It showed a 61.7% RMSE improvement and 57.1% MAPE improvement over the ANN model, and even more substantial gains of around 91.2% for RMSE and 90.91% for MAPE compared to the moving average method. Our long-term forecasting method is evaluated against alternative approaches, including ANN, moving average, and ANN + DWT. The ANN+DWT structure resembles ours but differs in input features, as it solely utilizes SOP change data without incorporating humidity and temperature. Results depicted in Tab. 2 show that our approach outperforms the other methods. Compared to ANN+DWT, our method reduces RMSE and MAPE by 41.8% and 46.3% respectively, underlining the importance of integrating tempera-ture and humidity for enhanced accuracy. Our approach sees around 53.64% and 52.03% RMSE and MAPE improvements. Furthermore, compared to the moving average method, our approach shows approximately 76.96% and 77.34% RMSE and MAPE enhancements, respectively.

Tab. 1: Prediction accuracy for short-term forecasting approaches: RMSE and MAPE

| Method | RMSE (rad/s) | MAPE (%) |
|---|---|---|
| Calm model | 1.28 | 0.6 |
| Windy model | 0.89 | 0.3 |
| ANN | 2.3 | 0.7 |
| Moving average | 10.12 | 3.3 |

Tab. 2: Prediction accuracy for long-term forecasting approaches: RMSE and MAPE.

| Method | RMSE (rad/s) | MAPE (%) |
|---|---|---|
| Proposed method | 7.1 | 2.9 |
| ANN + DWT | 12.2 | 5.4 |
| ANN | 15.1 | 6.15 |
| Moving average | 30.6 | 12.8 |

Fig. 4 depicts the SOP change forecasts generated by both short-term and long-term forecasting models from May 26 to May 31, 2021. The predicted values closely align with the true SOP change values, particularly for short-term predictions, indicating the effectiveness of our models in capturing the SOP dynamics within the specified timeframe. Enhancing long-term forecasting performance could be achieved by incorporating additional environmental factors impacting SOP change such as rainfall or snowfall.

**Conclusions**

Our approach combines neurowavelet modeling for multi-scale SOP change forecasting in aerial fiber links, integrating weather-related data. Validation with field data demonstrates its effectiveness for short-term and long-term forecasting. Additionally, incorporating environmental factors like wind gust, temperature, and humidity significantly enhances forecasting accuracy, underscoring their importance for precision.